\documentclass[prd,aps,preprint,amsmath,nofootinbib,amssymb,eqsecnum,showkeys,tightenlines]{revtex4}
\pdfoutput=1
\usepackage{verbatim,graphics,graphicx,color,slashed,textcomp}
\usepackage{ulem,soul,xcolor} 
\usepackage{hyperref}
\graphicspath{{figures/}}

\newcommand{\GeV}{\mathrm{GeV}}
\newcommand{\TeV}{\mathrm{TeV}}

\begin{document}
\setstcolor{red}

\title{Dark Higgs Channel for FERMI GeV $\gamma$-ray Excess }
\author{P. Ko and Yong Tang}
\affiliation{School of Physics, Korea Institute for Advanced Study,\\
 Seoul 130-722, South Korea }

\begin{abstract}
Dark Higgs is very generic in dark matter models where DM is stabilized by some spontaneously broken dark gauge symmetries. Motivated by tentative GeV scale $\gamma$-ray excess from the galactic center (GC), we investigate a scenario where a pair of dark matter $X$ annihilates into a pair of dark Higgs $H_2$, which subsequently decays into standard model particles through its mixing with SM Higgs boson. Besides the two-body decay of $H_2$, we also include multibody decay channels of the dark Higgs. We find that the best-fit point is around $M_X\simeq 95.0$GeV, $M_{H_2}\simeq 86.7$GeV, $\langle \sigma v\rangle\simeq 4.0\times 10^{-26}\textrm{cm}^3\textrm{/s}$ and gives a p-value $\simeq 0.40$.
Implication of this result is described in the context of dark matter models with  dark gauge symmetries. Since such a dark Higgs boson is very difficult to produce at colliders, indirect DM detections of cosmic $\gamma$-rays  could be an important probe of dark sectors, complementary to collider searches.  
\end{abstract}
\maketitle

\section{Introduction}
Firm evidences for dark matter (DM) come exclusively from the gravitational interaction at the moment. A popular scenario in particle physics models, DM as weakly-interacting massive particles (WIMP), generally predicts that DM should have a mass between $ \mathcal{O}(\GeV) - \mathcal{O}(\TeV)$, with a weak-scale  annihilation cross section around $3\times 10^{-26}\textrm{cm}^3\textrm{/s}$. If those annihilation final states go to standard model particles eventually, there might be notable excesses in cosmic rays and gamma ray searches.  

By analyzing \textit{Fermi-LAT}'s public data, several groups~\cite{Goodenough:2009gk, 
Hooper:2010mq, Boyarsky:2010dr, Hooper:2011ti, Abazajian:2012pn, Gordon:2013vta, Hooper:2013rwa,
Huang:2013pda, Abazajian:2014fta, Daylan:2014rsa, Calore:2014xka} have been 
claiming that there might be some excess in the gamma-ray signals from Galactic center~\footnote{Recently, Fermi-LAT also released an paper with some excess~\cite{TheFermi-LAT:2015kwa}.}, 
inner Galaxy and even some Dwarf Galaxy~\cite{Geringer-Sameth:2015lua}. 
The excess is at $E_\gamma \sim O($GeV) energy scale and its morphology against the distance to galaxy center is consistent with signals from WIMP DM annihilation, which has motivated intense discussions about DM model-constructions and constraints~\cite{Lacroix:2014eea, Alves:2014yha, Berlin:2014tja, Agrawal:2014una, Izaguirre:2014vva, Cerdeno:2014cda, Ipek:2014gua, Ko:2014gha, Boehm:2014bia, Abdullah:2014lla, Ghosh:2014pwa, Martin:2014sxa, Berlin:2014pya, Basak:2014sza, Cline:2014dwa, Han:2014nba, Detmold:2014qqa, Wang:2014elb, Chang:2014lxa, Arina:2014yna, Cheung:2014lqa, Huang:2014cla, Balazs:2014jla, Ko:2014loa, Baek:2014kna, Okada:2014usa, Bell:2014xta, Banik:2014eda, Borah:2014ska, Cahill-Rowley:2014ora, Yu:2014pra, Guo:2014gra, Cao:2014efa, Yu:2014mfa, Freytsis:2014sua, Heikinheimo:2014xza, Agrawal:2014oha, Cheung:2014tha, Arcadi:2014lta, Hooper:2014fda, Yuan:2014yda, Calore:2014nla, Liu:2014cma, 
Biswas:2014hoa, Ghorbani:2014gka, Cerdeno:2015ega, Biswas:2015sva, Alves:2015pea, Kaplinghat:2015gha, Berlin:2015sia, Chen:2015nea, Guo:2015lxa, Buckley:2015doa, 
Modak:2015uda, Caron:2015wda, Gherghetta:2015ysa, Elor:2015tva, Kopp:2015bfa, 
Bi:2015qva, Appelquist:2015yfa, Rajaraman:2015xka, Cline:2015qha, Kong:2014haa, Bringmann:2014lpa, Cirelli:2014lwa, Cholis:2014fja}. Besides DM interpretation, astrophysical origins of the excess have also been actively investigated~\cite{Yuan:2014rca, Carlson:2014cwa, Petrovic:2014uda, Fields:2014pia, Cholis:2014lta, Zhou:2014lva, Gordon:2014gya, O'Leary:2015gfa, Petrovic:2014xra, Bartels:2015aea, Lee:2015fea}. 
For instance, Refs.~\cite{Yuan:2014rca, Petrovic:2014xra} 
showed that gamma-ray emission from unresolved millisecond pulsars is compatible with the excess and can account for at least part of the excess. And Ref.~\cite{Bartels:2015aea, Lee:2015fea} argued that the excess might be comprised entirely of point sources. Recently, {\tt Fermi-LAT} collaboration published a paper~\cite{Ackermann:2015zua} on search for DM from Milky Way dwarf spheroidal galaxies and gave stringent constraints. However, since dwarf galaxies might have different DM density profiles from Milky Way halo, such constraints would be relaxed then. 

In this paper, we shall exclusively consider DM interpretations. As a first step, it is natural to investigate the GeV excess through annihilation channels that a pair of DM 
goes to two SM particles directly, such as $q\bar{q},c\bar{c},b\bar{b},t\bar{t},l^{\mp}l^{\pm},gg,hh,WW,ZZ$ 
and also their different combinations with some branching fractions. After all, no new particle has been found yet at the LHC, except the Higgs boson. This method has provided valuable information for the favored DM 
mass and annihilation cross section ranges. Discussions has been extended to cascade two-body decay 
through new mediators, such as $Z'$ and dark Higgs $H_2$, which are ubiquitous in new physics beyond SM. 
In particular, light mediator ($M_{Z',H_2}<1$GeV)~\cite{Liu:2014cma} and heavy $Z^{'}$~\cite{Cline:2014dwa} cases have been investigated thoroughly.

This work is intended to investigate GeV scale gamma-ray excess in models where DM 
annihilates into a pair of heavy dark Higgs ($>1$GeV) which subsequently could decay 
into multi-body final states such as $WW^*$ or $W^* W^*$, where $W^*$ is a virtual $W$ 
boson.   The aim is to provide the ranges of the favored dark Higgs mass, DM mass and 
the annihilation cross section, which might be useful for particle physics model building, 
such as hidden sector DM models with local dark gauge symmetries.  
This work differs from many previous investigation in one essential aspect: we take into 
account consistently all possible decay modes for heavy dark Higgs, not restricted to 
its two-body decays. 

This paper is organized as follows. In Sec.~\ref{sec:formalism}, we discuss the theoretical motivation and  establish our formalism and notations. In Sec.~\ref{sec:num}, we present our numerical results on the best-fit parameters.  Then in Sec.~\ref{sec:relic}, we briefly discuss 
the implications for DM relic density and constraints from our results.
Finally, we give a summary.

\section{Formalism}\label{sec:formalism}
We shall consider the following annihilation channel for self-conjugate DM $X$,
\begin{eqnarray*}
X & + & X  \rightarrow  H_2 + H_2,  \ \ \ {\rm followed ~by} 
 \\  H_2 & \rightarrow & SM + SM (+ SM).
\end{eqnarray*}
Here  $H_2$ denotes the dark Higgs, distinguishing it from the SM-like Higgs $H_1$ with 
$M_{H_1}\simeq 125$  GeV. In Ref.~\cite{Ko:2014loa}, the present authors showed that dark Higgs is very generic in dark matter models with dark gauge symmetries, and the GC 
$\gamma$-ray excess can be easily accommodated with the above mechanism (see also Refs.~\cite{Ko:2014gha,Boehm:2014bia,Baek:2014kna} for related works). 

$H_2$ can decay into SM particles through its small mixing with $H_1$. 
The mixture between $H_2$ and $H_2$ can be easily achieved in particle physics model building. 

For example, a real scalar DM $X$ and a complex scalar $\Phi$ (dark Higgs that breaks local dark gauge 
symmetries) can have the following interactions,
\begin{equation}\label{eq:lagrangian}
\mathcal{L}\supset -\lambda_{\phi X}X^2 \Phi^{\dagger}\Phi - \lambda_{\phi H}\Phi^{\dagger}\Phi H^{\dagger}H,
\end{equation}
where $H$ is the SM Higgs doublet. After gauge (or even possible global) symmetry breaking,
\begin{equation}
H\rightarrow \left(0,\frac{v_{h}+h}{\sqrt{2}}\right)^T,\;\phi\rightarrow\frac{v_{\phi}+\phi}{\sqrt{2}},
\end{equation}
where $v_h$ and $v_\phi$ are the vacuum expectation values, two neutral scalars $h$ and $\phi$ will mix 
with each other through the Higgs portal coupling $\lambda_{\phi H}$, resulting in two mass eigenstates 
$H_{1}$ and $H_{2}$ with
\begin{equation}\label{eq:mixingangle}
\left(\begin{array}{c}
H_{1}\\
H_{2}
\end{array}\right)=\left(\begin{array}{cc}
\cos{\alpha} & {}-\sin{\alpha}\\
\sin{\alpha} & \cos{\alpha}
\end{array}\right)\left(\begin{array}{c}
h\\
\phi
\end{array}\right),
\end{equation}
in terms of the mixing angle $\alpha$. 

The above Lagrangian is just one example of many DM models with Higgs portal. One can also consider 
the case with a real scalar $\phi$, 
\begin{equation}\label{eq:lgrg2}
\mathcal{L}\supset -\lambda_1 X^2 \phi -\lambda_2 X^2 \phi ^2 - \lambda_3\phi H^{\dagger}H - \lambda_4\phi^2 H^{\dagger}H.
\end{equation}
Again after the electroweak symmetry breaking,
\begin{equation}
H\rightarrow \left(0,\frac{v_{h}+h}{\sqrt{2}}\right)^T,
\end{equation}
$h$ and $\phi$ are mixed. We can also consider a model with fermionic $X$ with 
\begin{equation}\label{eq:lgrg3}
\mathcal{L}\supset -\lambda_1 \bar{X}\gamma _5 X \phi - \lambda_2\phi H^{\dagger}H - \lambda_3\phi^2 H^{\dagger}H.
\end{equation}
All the above models can easily evade current experimental bounds when $\alpha$ is very small 
(see discussion in Ref.~\cite{Baek:2014goa} for example). If one assumes all other possible new particles are 
heavy, DM $X$ will dominantly annihilate into $H_2$'s. 

\begin{figure}[t]
\includegraphics[width=0.35\textwidth, height=0.20\textwidth]{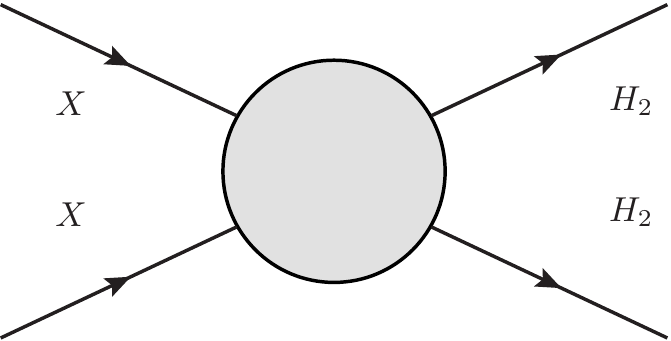}
\caption{Feynman diagram due to the effective operator $X^2H^2_2$ ($\bar{X}\gamma _5X H^2_2$ for fermionic $X$ or $X_\mu X^\mu H^2_2$ for vector $X$). The actual annihilation process may occur through s or t channel, and contact interaction. Details in the gray bubble depend on various ultraviolet completions. The produced $H_2$s can have two-, three- or even four-body decay channels. \label{fig:diagram}}
\end{figure}

To be as general as possible, we shall just work with the effective operator, $X^2 H^2_2$ 
($\bar{X}\gamma _5X H^2_2$ for fermionic $X$), and consider the annihilation process in 
Fig.~\ref{fig:diagram}, assuming that other particles in the dark sector are all heavy enough.

The produced $H_2$'s could be either relativistic or non-relativistic for $M_{H_2}\ll M_X$ or 
$M_{H_2}\simeq M_X$, respectively. Different kinematics might lead to significant differences in the 
gamma-ray spectra. Moreover, depending on the mass of $H_2$, $M_{H_2}$, $H_2$ can dominantly decay 
into 2 or 3 standard model particles. In our numerical calculation, we use \texttt{PYTHIA-6.4}~\cite{Pythia} to 
simulate and tabulate $d N^f_\gamma/dE_\gamma$ for the interesting ranges of $M_X$ and $M_{H_2}$. 
In particular, we focus on $M_X \geq 5$ GeV and $M_{H_2}\geq 1$ GeV.

The general differential flux of the gamma-ray  from the annihilation of self-conjugate DM is given by
\begin{equation}\label{eq:flux}
\frac{d^2 \Phi}{dE_\gamma d\Omega}=\frac{1}{8\pi}\sum_f\frac{\langle \sigma v\rangle^f_\textrm{ann}}{ M_{\textrm{DM}}^2}\frac{d N^f_\gamma }{dE_\gamma}\int_0^\infty dr \rho ^2 \left(r'\left( r,\theta \right) \right),
\end{equation}
where $\langle \sigma v\rangle^f_\textrm{ann}$ is the velocity-averaged annihilation cross section for the annihilation channel $f$, $d N^f_\gamma/dE_\gamma$ is prompt gamma-ray spectrum, $r'=\sqrt{r_\odot^2 + r^2 -2r_\odot  r \cos \theta}$, $r$ is the distance to earth from the DM annihilation point,  $r_\odot\simeq 8.5$kpc for solar system and $\theta$ is the observation angle between the line-of-sight and the center of Milky Way. An extra factor $1/2$ needs to be included for non-self-conjugate DM, such as complex scalars or Dirac fermions. In our considered case, we have only one annihilation channel, $X + X \rightarrow H_2 + H_2$.

For DM density distribution, we use the following generalized NFW profile~\cite{Navarro:1995iw},
\begin{equation}\label{eq:haloprofile}
\rho\left(r\right)=\rho_\odot \left[\frac{r_\odot}{r}\right]^\gamma \left[\frac{1+r_\odot/r_c}{1+r/r_c}\right]^{3-\gamma},
\end{equation}
with parameters $r_c\simeq 20$kpc and $\rho_\odot \simeq 0.4\textrm{GeV}/\textrm{cm}^3$. 
We shall adopt the index $\gamma=1.26$ if not stated otherwise. 

\section{Numerical Analysis}\label{sec:num}
\begin{figure}[t]
\includegraphics[width=0.65\textwidth, height=0.58\textwidth]{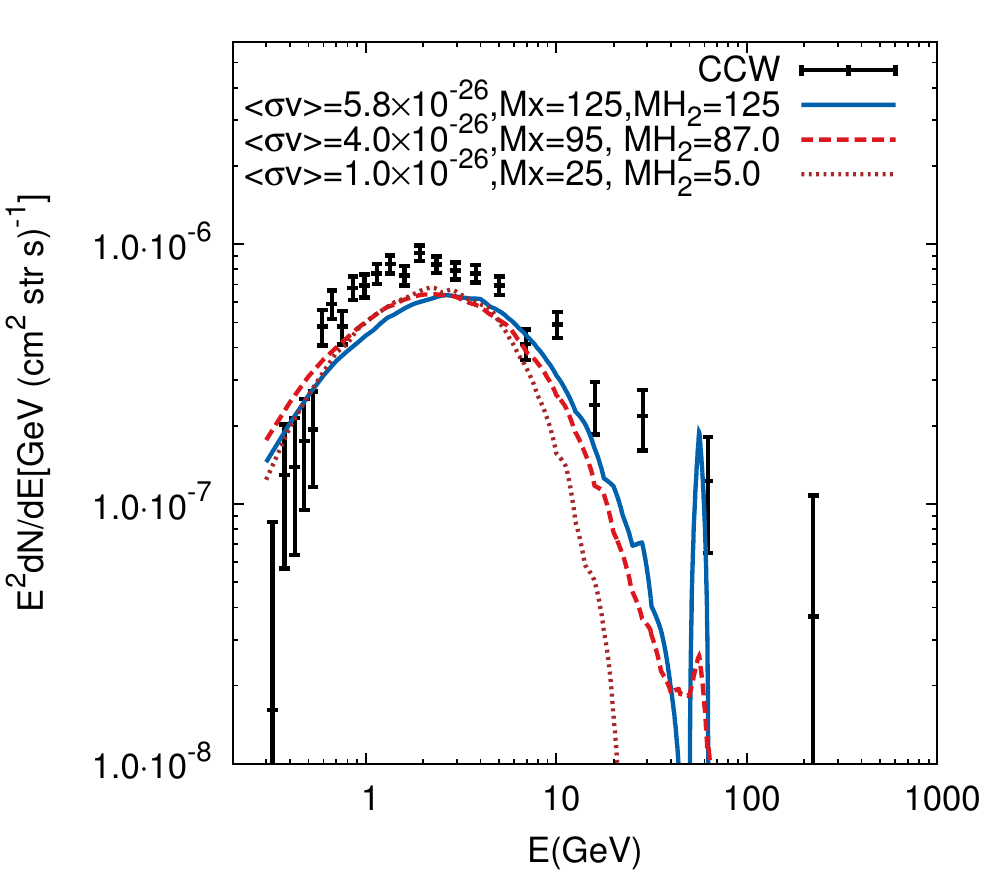}
\caption{Three illustrative cases for gamma-ray spectra in contrast with CCW data points~\cite{Calore:2014xka}. All masses are in GeV unit and $\sigma v$ with cm$^3$/s. Line shape around $E\simeq M_{H_2}/2$ is due to decay modes, $H_2\rightarrow \gamma\gamma, Z\gamma$.\label{fig:gamma}}
\end{figure}
We first show three cases for the gamma-ray spectrum in Fig.~\ref{fig:gamma}. The vertical axis marks the conventional
\begin{equation}
E^2\frac{dN}{dE}\equiv E_\gamma^2 \frac{1}{\Delta \Omega}\int _{\Delta \Omega} \frac{d^2 \Phi}{dE_\gamma d\Omega},
\end{equation}
where $\Delta \Omega$ indicates the region of interest. The 24 data points we used to compare with are from Ref.~\cite{Calore:2014xka}, denoted as CCW hereafter.

As we can see, different parameter sets can give different spectrum shape, especially in the high energy regime. When the branching ratios of $H_2\rightarrow \gamma \gamma, Z\gamma$ are increasing, we can see the gamma lines more easily around $E\simeq M_{H_2}/2$. Since the annihilation cross section is at order of $10^{-26}$cm$^3$/s and the branching ratios of $H_2\rightarrow \gamma \gamma, Z\gamma$ are around $0.2\%$ at most, the considered parameters are still consistent with constraint from gamma-line searches.

We now use the $\chi^2$ function and find its minimum to find out the best fit:
\begin{equation}
\chi^2\left(M_X, M_{H_2},\langle \sigma v\rangle\right)=\sum_{i,j}\left(\mu_i - f_i \right)
\Sigma^{-1}_{ij}\left(\mu_j - f_j \right),
\end{equation}
where $\mu_i$ and $f_i$ are the predicted and measured fluxes in the $i$-th energy bin respectively, 
and $\Sigma$ is the $24\times 24$ covariance matrix. We take the numerical values for $f_i$ and $\Sigma$ from CCW~\cite{Calore:2014xka}. 
Minimizing the $\chi^2$ against $f_i$ with respect to $M_X$, $M_{H_2}$ and $\langle \sigma v \rangle$ gives 
the best-fit points, and then two-dimensional $1\sigma$, $2\sigma$ and $3\sigma$ contours are defined 
at $\Delta \chi^2\equiv \chi^2-\chi^2_{\textrm{min}}=2.3$, $6.2$ and $11.8$, respectively.

\begin{figure}[t]
\includegraphics[width=0.49\textwidth, height=0.49\textwidth]{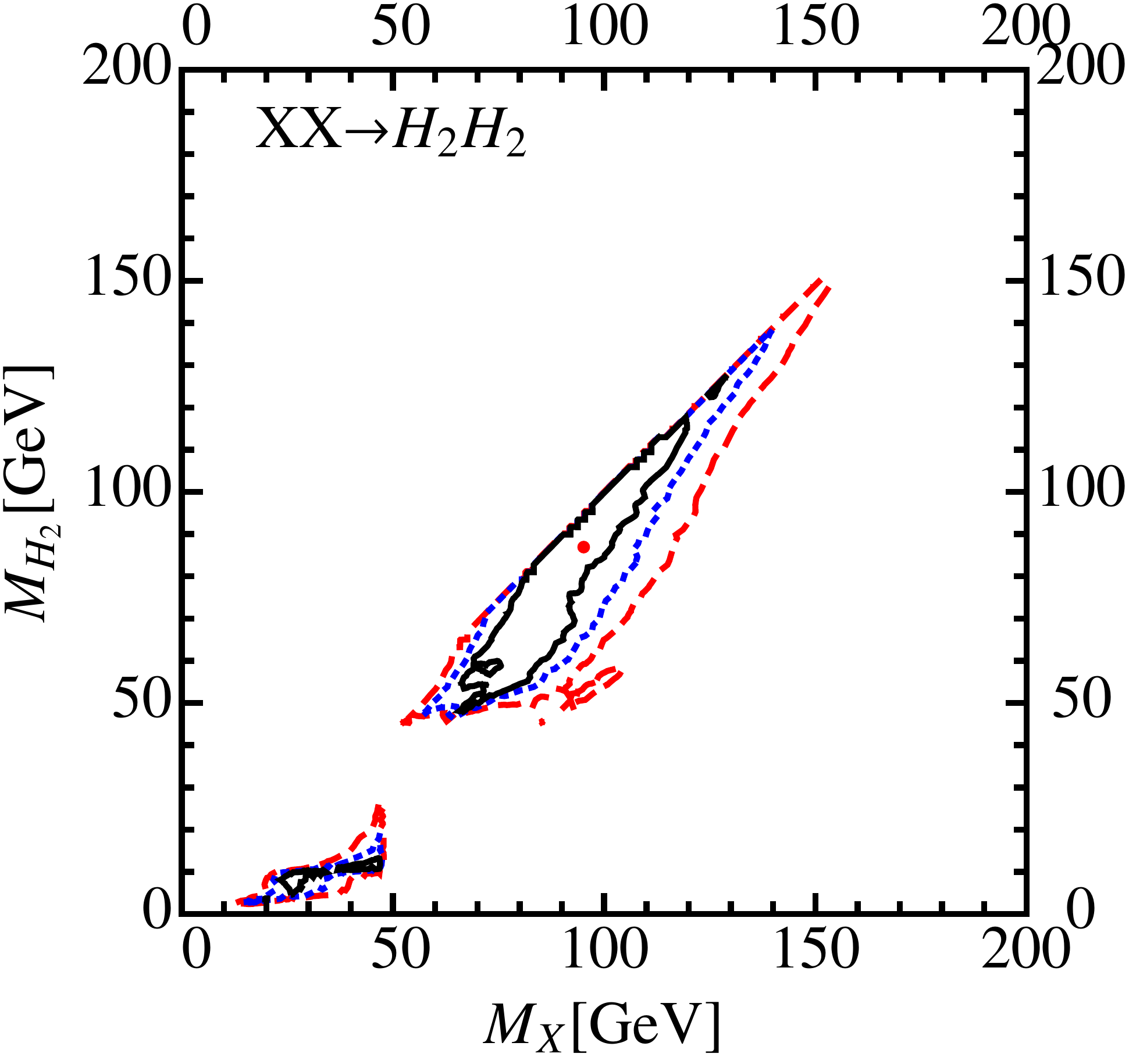} 
\includegraphics[width=0.49\textwidth, height=0.49\textwidth]{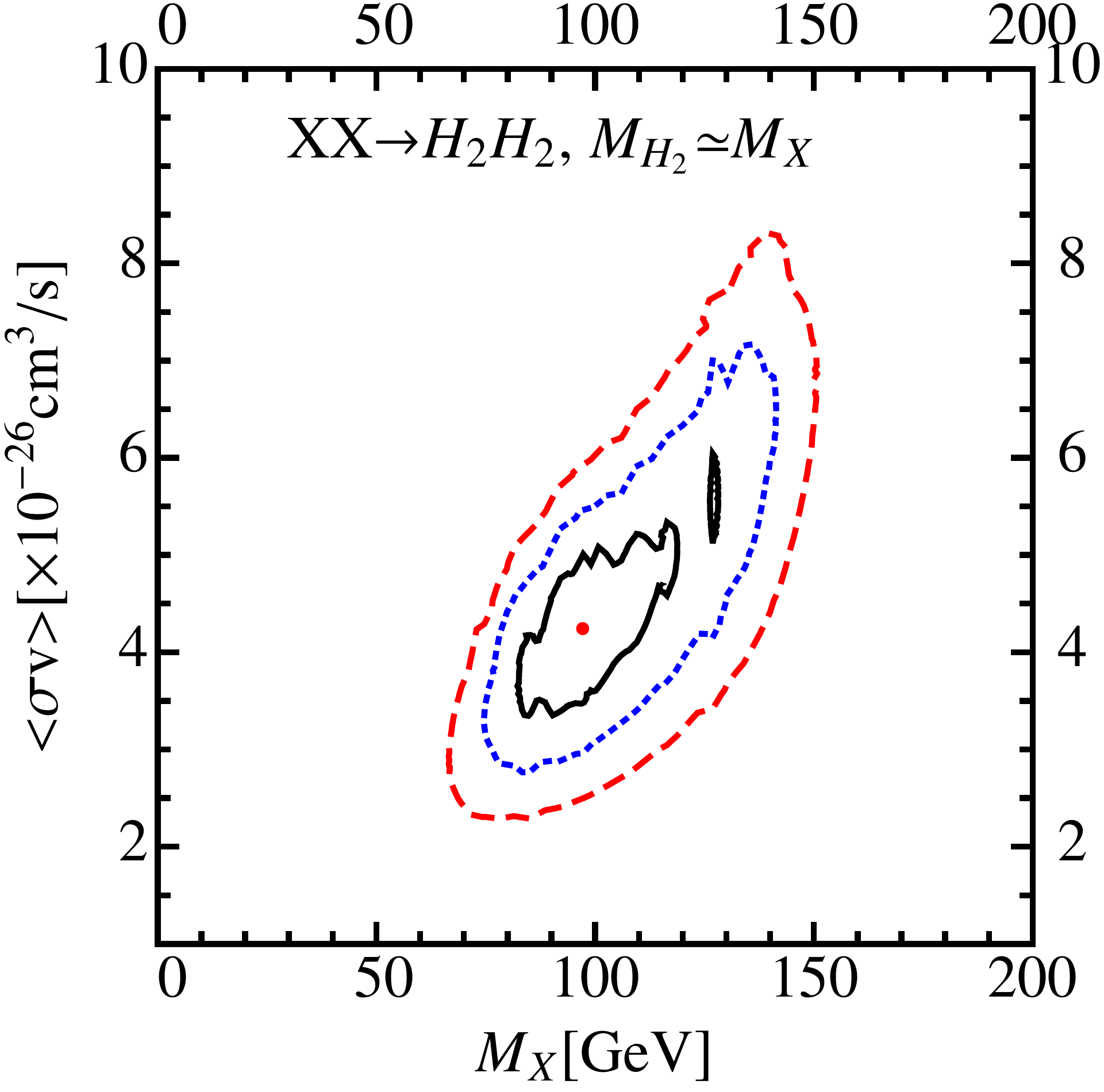}
\caption{The regions inside solid(black), dashed(blue) and long-dashed(red) contours correspond to $1\sigma$, $2\sigma$ and $3\sigma$, respectively. The red dots inside 1$\sigma$ contours are the best-fit points. 
In the left panel, we vary freely $M_X$, $M_{H_2}$ and $\langle \sigma v\rangle$. While in the right panel, we fix the mass of $H_2$, $M_{H_2}\simeq M_X$.
\label{fig:fits}}
\end{figure}
\begin{figure}[htb]
\includegraphics[width=0.49\textwidth, height=0.49\textwidth]{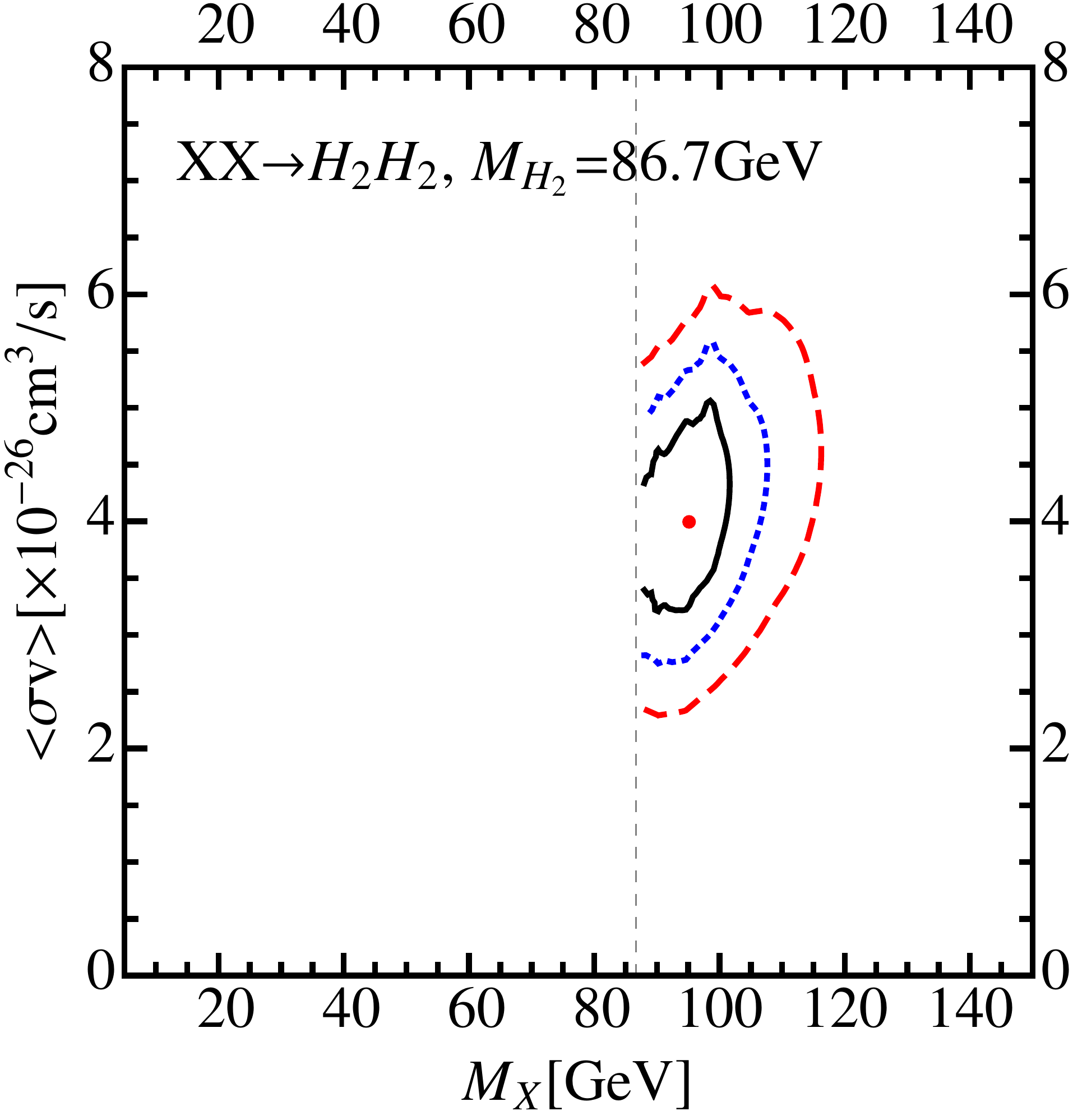} 
\includegraphics[width=0.49\textwidth, height=0.49\textwidth]{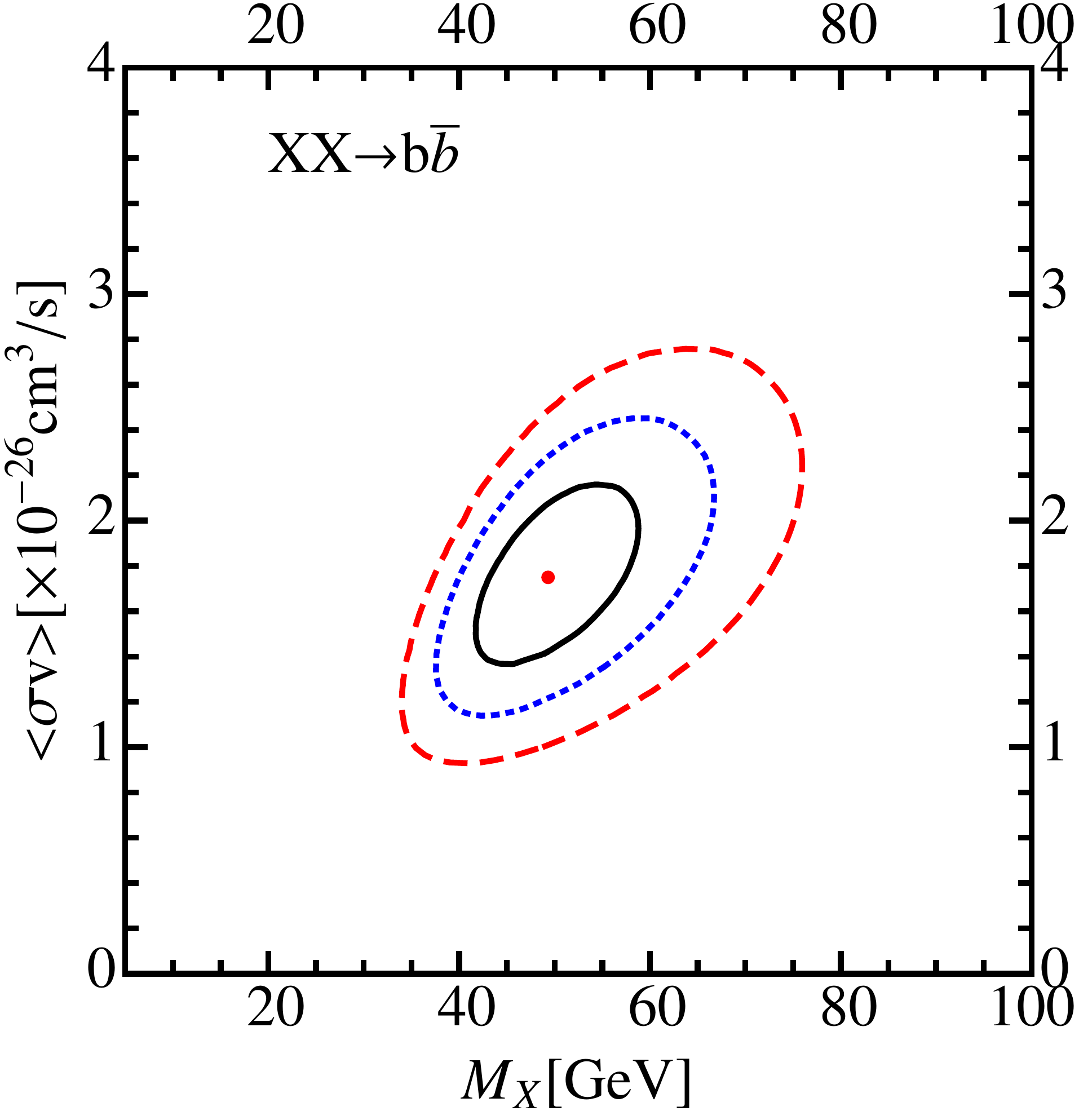}
\caption{Regions inside solid(black), dashed(blue) and long-dashed(red) contours correspond to $1\sigma$, $2\sigma$ and $3\sigma$, respectively. The red dots inside 1$\sigma$ contours are the best-fit points. In the left panel, we fix $M_{H_2}=86.7$GeV, but vary freely $M_X$ and $\langle \sigma v\rangle$. For comparison, in the right panel we consider $XX\rightarrow b\bar{b}$ channel. 
\label{fig:bbhh}}
\end{figure}

Fig.~\ref{fig:fits} is our main result. In the left panel, $M_X$, $M_{H_2}$ and $\langle \sigma v\rangle$ are  
freely varied, so that the total degree of freedom (d.o.f.) is $21$. The red dot represents the best-fit point with  
\begin{equation}\label{eq:hh1}
M_X\simeq 95.0\GeV,\; M_{H_2}\simeq 86.7\GeV,\;\langle \sigma v\rangle\simeq 4.0\times  10^{-26}\textrm{cm}^3\textrm{/s},  
\end{equation}
gives $\chi^2_{\textrm{min}}\simeq 22.0$,  with the corresponding p-value equal to $0.40$. 

We also notice that there are two separate regimes, one in the low mass region and the other in high mass 
region. 
The higher mass region is basically aligned with $M_{H_2}\simeq M_X$ since otherwise a highly-boosted  
$H_2$ would give a harder gamma-ray spectrum. In this region, $H_2$ mostly decays into $b\bar{b}$. As one increases the mass of $H_2$, $H_2\rightarrow W^{\pm}l^{\mp}\nu$, $H_2\rightarrow Z l^{\pm}l^{\mp}$, $H_2\rightarrow \gamma \gamma$ and $H_2\rightarrow \gamma Z$ become more and more important, and 
all of them give harder gamma-ray spectra either due to the leptonic final states or the gamma lines. 
Eventually, $\chi^2$ increases significantly when $M_{H_2}\geq 150\GeV$.  

In the low mass region, the contours are scattered but centered around $M_{H_2}\simeq 10$GeV and such 
a light $H_2$ most likely decays into $b\bar{b},c\bar{c}$ and $\tau^+\tau^-$. Since $c\bar{c}$ and $\tau^+\tau^-$ would give harder spectra than $b\bar{b}$ does, we would need a lower $M_X$ to fit the data, which is exactly 
what we see in Fig.~\ref{fig:gamma} (dotted curve). Increasing the branching ratios of $c\bar{c}$ and 
$\tau^+\tau^-$ would require a even lower $M_X$. 

In the right panel, we show a special case in which $M_{H_2}\simeq M_X$, so that the produced $H_2$s are non-relativistic. In such a case, the d.o.f. is now 22. The best-fit parameters are
\begin{equation}\label{eq:hh2}
M_X\simeq  M_{H_2}\simeq  97.1\GeV,\;\langle \sigma v\rangle\simeq 4.2\times  10^{-26}\textrm{cm}^3\textrm{/s}, 
\end{equation}
which gives $\chi^2_{\textrm{min}}\simeq 22.5$ and the p-value equal to $0.43$. An interesting thing is that 
$M_X\simeq M_{H_2}\simeq 125$GeV also give a good-fit. This point is equivalent to the channel that DM 
$X$ annihilates into SM Higgs, which has been already found in previous study 
\cite{Agrawal:2014oha,Calore:2014nla}. 

In the left panel of Fig. ~\ref{fig:bbhh}, we fix the mass of dark Higgs to the best-point value, $M_{H_2}=86.7$, and vary $M_X$ and $\langle \sigma v\rangle$. We show $1\sigma$, $2\sigma$ and $3\sigma$ contours in terms of solid(black), dashed(blue) and long-dashed(red) curves, respectively. To compare with $b\bar{b}$ channel, we also present $3\sigma$ region in the right panel of Fig.~\ref{fig:bbhh}. The best-fit point is around
\begin{equation}\label{eq:bb}
M_X\simeq 49.4\GeV,\; \langle \sigma v\rangle\simeq 1.75 \times  10^{-26}\textrm{cm}^3\textrm{/s}, 
\end{equation}
which gives $\chi^2_{\textrm{min}}\simeq 24.4$ and a p-value, $0.34$.

\begin{table}[t]
  \centering
    \begin{tabular}{|c||c|c|c|}
    \hline 
Channels  &  Best-fit parameters & $\chi^2_{\textrm{min}}$/d.o.f. & $p$-value \\
  \hline
  $XX\rightarrow H_2 H_2$ & $M_X\simeq 95.0\GeV, M_{H_2}\simeq 86.7\GeV$ & 22.0/21 & 0.40 \\
(with $M_{H_2} \neq M_X$)  &  
$\langle \sigma v\rangle\simeq 4.0 \times  10^{-26}\textrm{cm}^3\textrm{/s}$ & & \\   
  \hline
  $XX\rightarrow H_2 H_2$ & $M_X\simeq 97.1\GeV$  & 22.5/22 & 0.43 \\
 (with $M_{H_2} = M_X$) & $\langle \sigma v\rangle\simeq 4.2 \times  10^{-26}\textrm{cm}^3\textrm{/s}$
 & &   \\
   \hline
  $XX\rightarrow H_1 H_1$ & $M_X\simeq 125\GeV$  & 24.8/22 & 0.30 \\
 (with $M_{H_1} = 125\GeV$) & $\langle \sigma v\rangle\simeq 5.5 \times  10^{-26}\textrm{cm}^3\textrm{/s}$
 & &   \\ 
  \hline
  $XX\rightarrow b\bar{b}$ & $M_X\simeq 49.4\GeV$  & 24.4/22 & 0.34 \\
  & $\langle \sigma v\rangle\simeq 1.75 \times  10^{-26}\textrm{cm}^3\textrm{/s}$ & & \\
  \hline 
    \end{tabular}
 \caption{Summary table for the best fits with three different assumptions. \label{tab:summary}}
\end{table}

\begin{figure}[t]
\includegraphics[width=0.49\textwidth, height=0.49\textwidth]{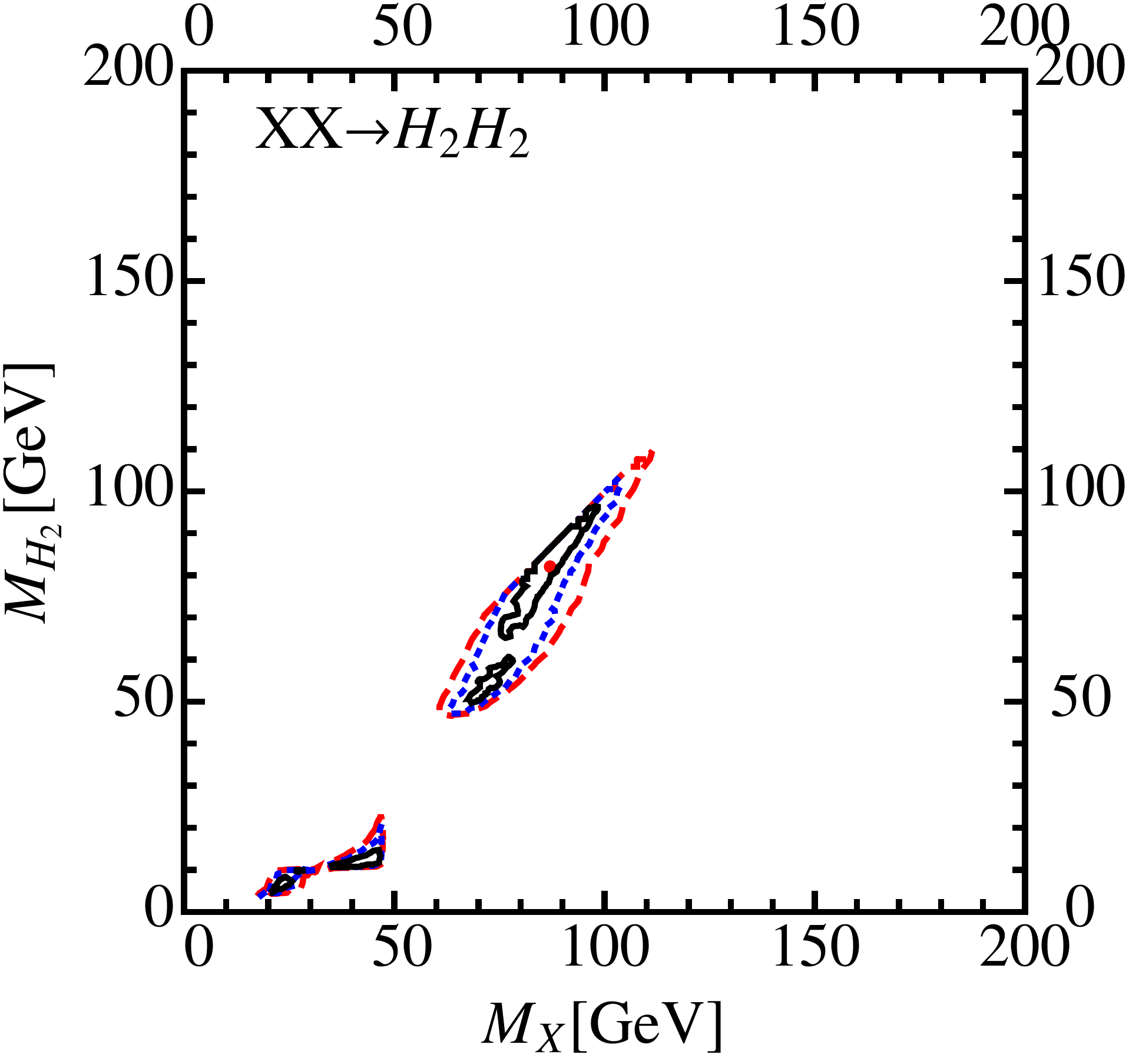} 
\includegraphics[width=0.49\textwidth, height=0.49\textwidth]{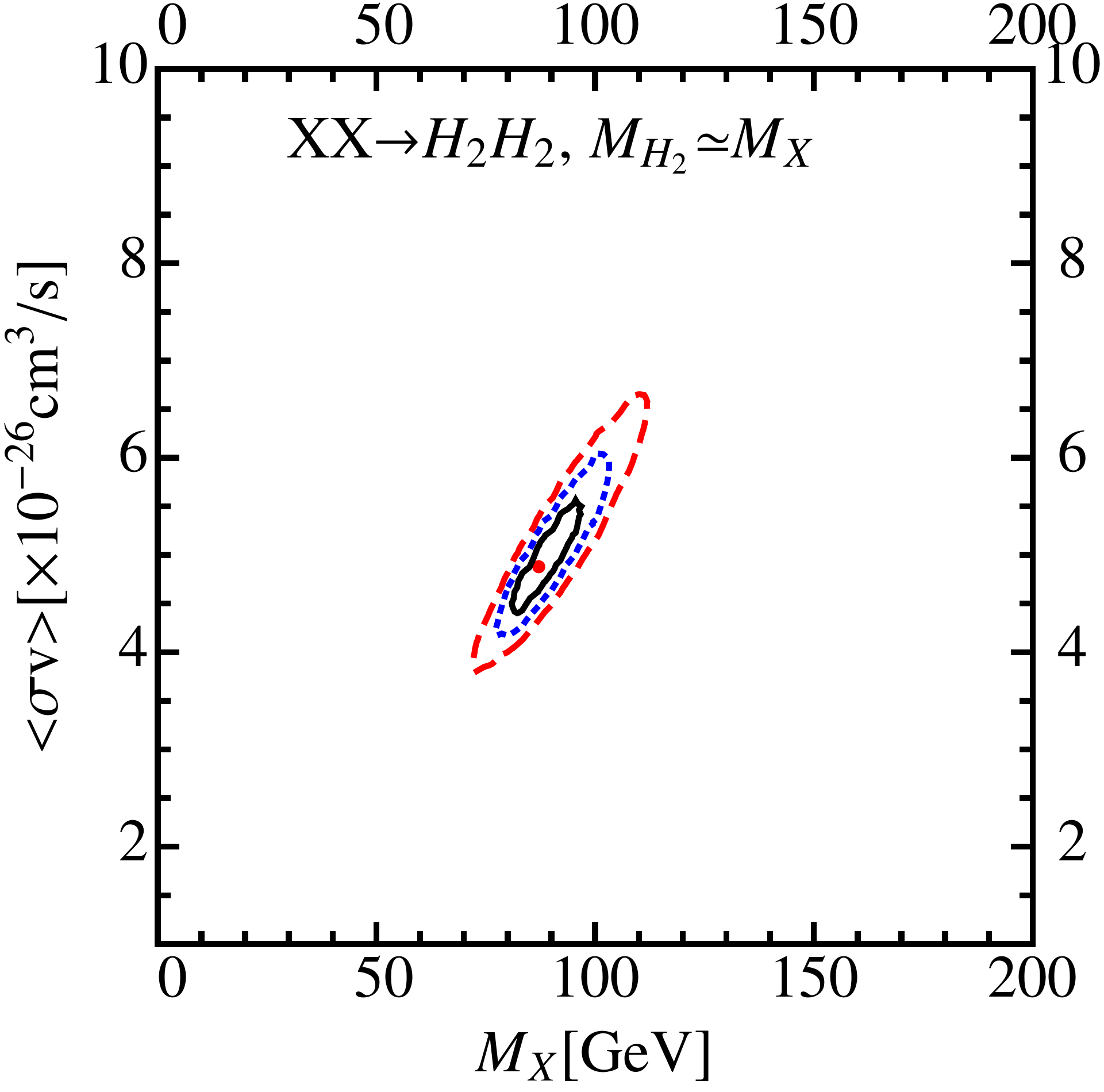}
\includegraphics[width=0.49\textwidth, height=0.49\textwidth]{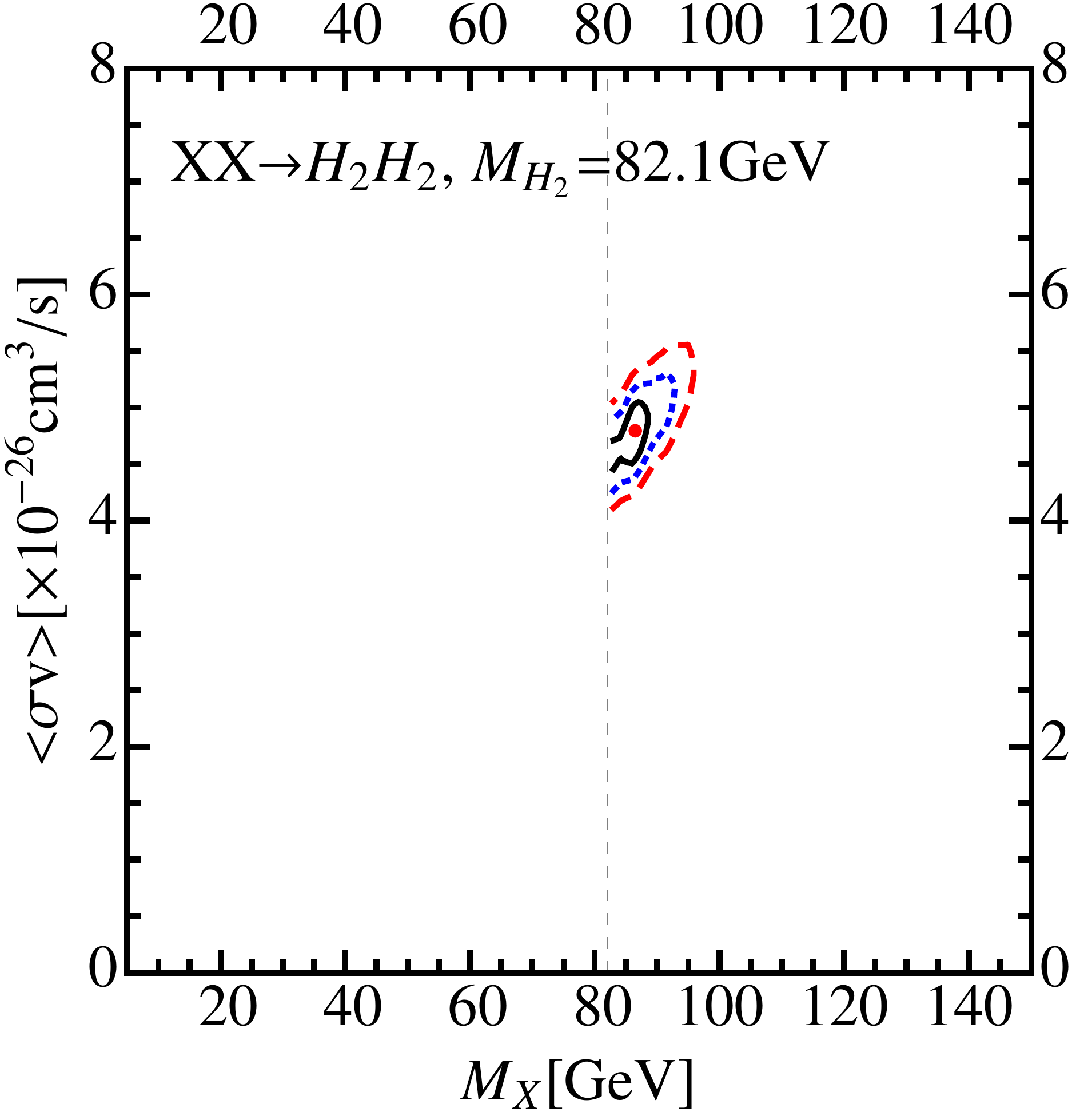} 
\includegraphics[width=0.49\textwidth, height=0.49\textwidth]{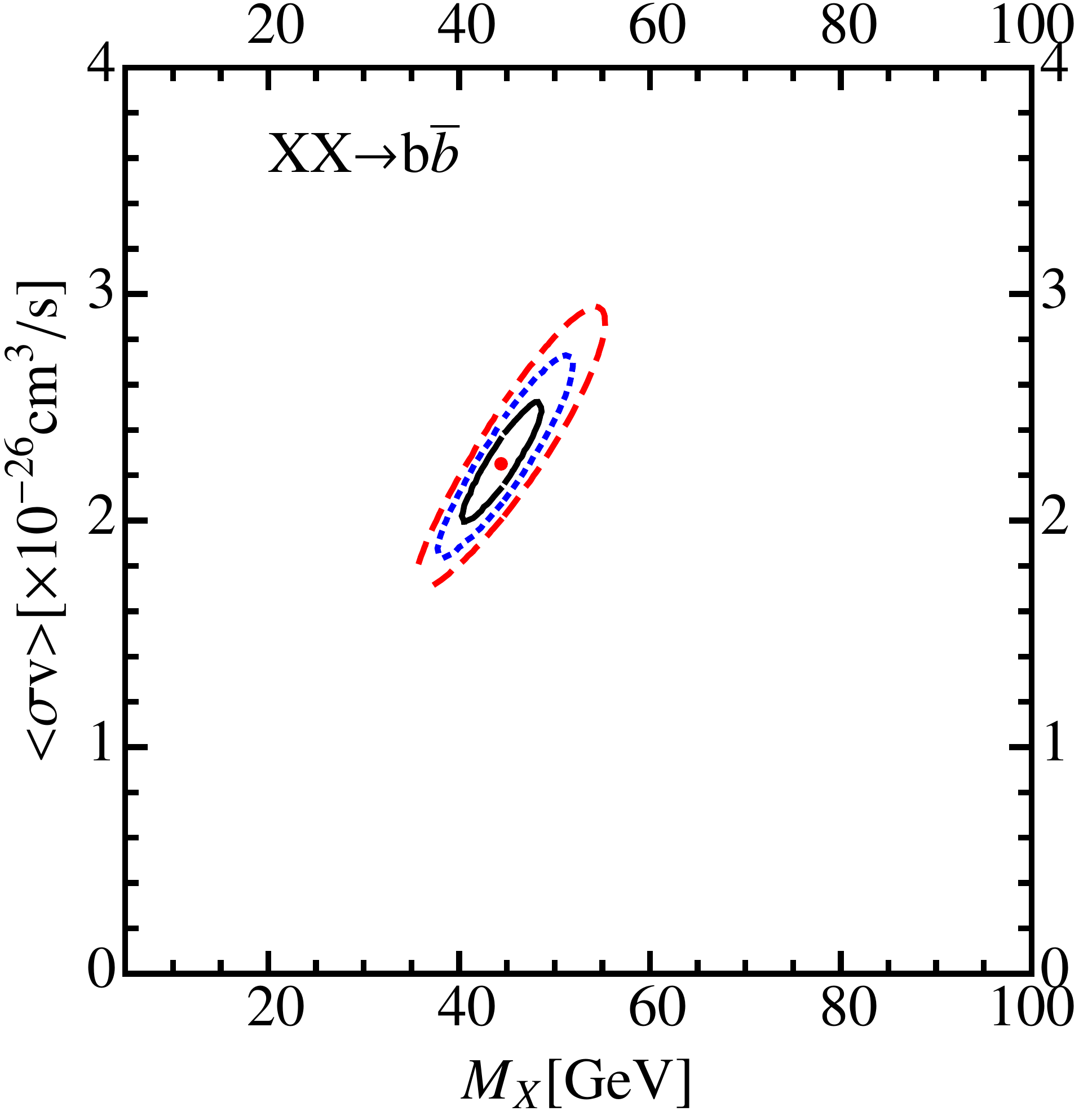}
\caption{Similar plots as Figs.~\ref{fig:fits} and \ref{fig:bbhh}, based on results from Ref.~\cite{Daylan:2014rsa}. Regions inside solid(black), dashed(blue) and long-dashed(red) contours correspond to $1\sigma$, $2\sigma$ and $3\sigma$, respectively. The red dots inside 1$\sigma$ contours are the best-fit points.
\label{fig:fits_2}}
\end{figure}

For completeness and comparison, we also show similar plots in Fig.~\ref{fig:fits_2} based on the analysis of Inner Galaxy from Ref.~\cite{Daylan:2014rsa}. All parameters are the same as previous except $\gamma=1.18$ for the DM profile as used Ref.~\cite{Daylan:2014rsa}. As shown in Fig.~\ref{fig:fits_2}, the regions inside contours are much smaller than those in Figs.~\ref{fig:fits} and \ref{fig:bbhh}. This is due to the fact that uncertainties in Ref.~\cite{Daylan:2014rsa} are purely statistical. All the best-fit points give $\chi^2\sim 44$, and the minimal $\chi^2$ is reached when $M_X\simeq 87.0\GeV$, $M_{H_2}\simeq 82.1\GeV$ and $\langle \sigma v\rangle\simeq 4.7 \times  10^{-26}\textrm{cm}^3\textrm{/s}$ with $\chi^2_{\textrm{min}}/\textrm{d.o.f} \simeq 42.6/(22-3)$ which corresponds to a p-value, $1.5\times 10^{-3}$. It seems that this is not a good fit. However, according to Ref.~\cite{Daylan:2014rsa}, any value of $\chi^2 \lesssim 50$ should be taken as a reasonable ``good fit'', given the large systematic uncertainties associated with the background templates choice. Nevertheless, we can easily see that the favored regions are consistent with those in Figs.~\ref{fig:fits} and \ref{fig:bbhh} within $2$-$3\sigma$. 

\section{Relic Abundance and Phenomenology}\label{sec:relic}
So far, we have not discussed any actual concrete DM models, how the correct relic abundance can be achieved and how one can test or constrain this scenario further. 
Generally, such topics are highly model dependent and the conclusions would differ 
significantly from one case to another.  Still, we can give some general implications 
from the results obtained earlier in this letter.

From Eqs.~(\ref{eq:hh1}) and (\ref{eq:hh2}), we notice that the best-fit annihilation cross 
section for dark Higgs channel is bigger than the canonical value for thermal DM $X$, 
$2\sim 3\times  10^{-26} \textrm{cm}^3 \textrm{/s}$, although still consistent within 
$2$-$3\sigma$ range.   Larger $\langle \sigma v \rangle$ would mean a smaller relic 
abundance for $X$.   However, the correct relic density can still be reached if we extend 
the above minimal DM model setup. For example, we may introduce another heavy 
DM component $Y$ and the total energy fraction of $\Omega_Y + \Omega_X$ is just the 
observed $\Omega_{\textrm{DM}}$.  Suppose $Y$ froze out in the early universe but had 
decayed into $X$ pairs with a lifetime shorter than the age of Universe. This can be easily achieved if we introduce the following interactions:
\[
\delta \mathcal{L} = y_1 Y^2 X^2 + y_2 M_X Y X^2.
\]
When $y_2$ is small enough, say less than $10^{-12}$, $Y$ would decay 
into $X$ pair after BBN epoch or the  freeze-out of $X$'s.  Before its decay, $Y$'s relic abundance was determined by $y_1$ through the efficient annihilation process, $Y + Y\rightarrow X + X$.  Although $y_2$ is small, it is still technically natural or natural by 't Hooft's naturalness argument because if $y_2$ is zero, we have an additional $Z_2$ symmetry for particle $Y$.

For $b\bar{b}$ channel, the best-fit cross section, Eq.~(\ref{eq:bb}), is a little smaller than the canonical value. This is not a problem at this stage since the right panel of Fig.~\ref{fig:bbhh} shows that canonical value of thermal cross section is within $1$-$2\sigma$ range. Even if it is a problem, we can still imagine that in some concrete models, there could exist some other annihilation channels whose cross sections have velocity dependence or $p$-wave suppression, $\langle \sigma v\rangle \propto v^2$. For instance, if fermionic DM $X$ has interaction, $\bar{X}X\phi^2$, then $X+\bar{X}\rightarrow 2\phi$ is much more suppressed nowadays than at the freeze-out time. Then the sum of cross sections to $\phi \phi$ and $b\bar{b}$ has a canonical value at the freeze-out time but only $b\bar{b}$ channel is important now for its indirect detections. 

Here we briefly discuss the related possible particle phenomenology in an explicit UV complete model. We shall note phenomenology is highly model-dependent and we refer to our previous works~\cite{Baek:2014goa, Ko:2014gha, Ko:2014loa, Baek:2014kna} for some detailed examples. Our discussion here is just focused on the following Lagrangian with $U_X(1)$ gauge symmetry:
\begin{eqnarray}
\mathcal{L}& = & \mathcal{L}_{\mathrm{SM}} - \frac{1}{4} X_{\mu\nu} X^{\mu\nu} +  
(D_\mu \Phi)^\dagger (D^\mu \Phi) 
- \lambda_\Phi \left( \Phi^\dagger  \Phi - v_\Phi^2/2 \right)^2
\nonumber \\
& & - \lambda_{\Phi H} \left(\Phi^\dagger \Phi - v_\Phi^2/2\right) 
\left(H^\dagger H - v_H^2/2\right) \ ,
\label{eq:full_theory}
\end{eqnarray}
where in the SM Lagrangian the Higgs potential term is $ \lambda_H \left( H^\dagger  H - v_H^2/2 \right)^2$ and the covariant derivative is defined as $ D_\mu \Phi = (\partial_\mu - i g_X X_\mu) \Phi$.

A nonzero vacuum expectation value $v_\Phi$ breaks $U(1)_X$ spontaneously. Afterwards, $X_\mu$ gets mass $M_X = g_X v_\Phi$ and the dark Higgs field $\phi$ will mix with the SM Higgs field $h$ through the Higgs portal term, resulting in two mass eigenstates, $H_1$ and $H_2$. The mixing angle $\alpha$ in the matrix Eq.~(\ref{eq:mixingangle}) is determined by
\begin{eqnarray}
\sin2\alpha=\frac{2\lambda_{H\Phi}v_{H}v_{\Phi}}{m_{H_2}^{2}-m_{H_1}^{2}}.
\end{eqnarray}
In this model, gauge boson $X_\mu$ is the dark matter because its only couplings, $X_\mu X^\mu H^2_{1,2}$ and $X_\mu X^\mu H_{1,2}$, have accidental $Z_2$ symmetry which ensure the stability.

The mixing angle is constrained by Higgs signal strength and invisible decay at the LHC~\cite{Cheung:2015dta, CMS:2015naa}, 
\[
\sin^2\alpha \lesssim 0.2, \textrm{ and } \lambda_{H\Phi}\lesssim 10^{-2} \textrm{ for } M_{H_2}<62.5\GeV.
\] 

\begin{figure}[tb]
\includegraphics[width=1\textwidth]{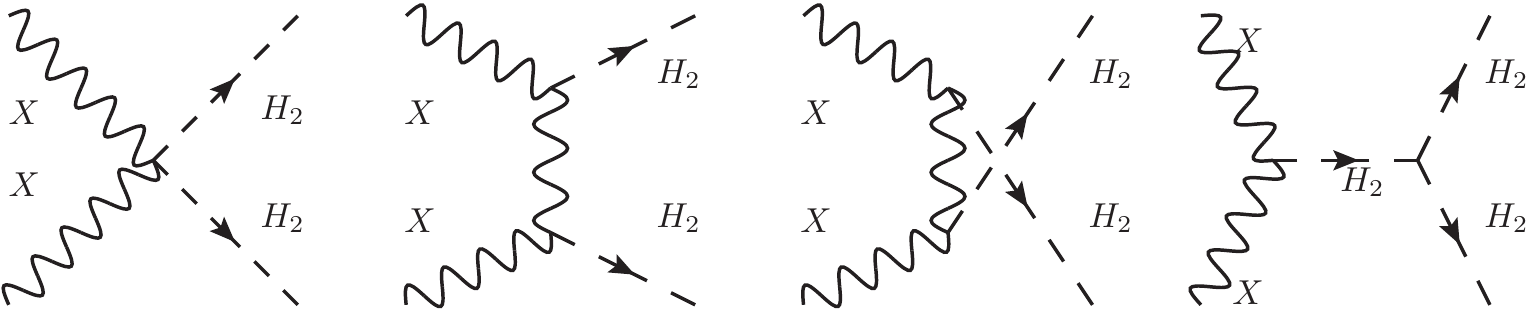}
\caption{Feynman diagrams. The last one can be neglected for small $\lambda_{\Phi}$. \label{fig:diagrams} }
\end{figure}

Due to the allowed small mixing, the dominant annihilation processes for small scalar self-coupling are shown in Fig.~\ref{fig:diagrams}, their thermal cross section $\langle\sigma v\rangle$ to determine the relic density is given by
\begin{eqnarray}\label{eq:sigmav}
\langle\sigma v\rangle = \frac{g_{X}^{4}\cos^4\alpha}{144\pi M_{X}^{2}} \left[3-\frac{8\left(M_{H_{2}}^{2}-4M_{X}^{2}\right)}{M_{H_{2}}^{2}-2M_{X}^{2}}
+ \frac{16\left(M_{H_{2}}^{4}-4M_{H_{2}}^{2}M_{X}^{2}+6M_{X}^{4}\right)} {\left(M_{H_{2}}^{2}-2M_{X}^{2}\right)^{2}}\right].
\end{eqnarray}
For the above formula, we can determine the required gauge coupling $g_X$ for correct DM density. For instance, if $M_{H_2}\ll M_{X}$, $\langle\sigma v\rangle\sim 3\times 10^{-26}\mathrm{cm^3/s}$ gives 
\[g_X \sim \frac{0.2}{\cos\alpha}\left(\frac{M_X}{100\GeV}\right)^{1/2}.\] 
While if $M_{H_2}\simeq M_{X}$, $\langle\sigma v\rangle\sim 3\times
10^{-26}\mathrm{cm^3/s}$ would give
\[g_X \sim \frac{0.24}{\cos\alpha}\left(\frac{M_X}{100\GeV}\right)^{1/2}.\]

The scalar mixing also lead to possible signal for direct detection of DM. The spin-independent scattering cross section for exchanging scalar mediators is calculated as
\begin{eqnarray}
\sigma_p^{\rm SI} \label{sp-th}
&=& \frac{g^2_X m^2_p f_p^2\sin^2 2\alpha }{4\pi v^2_H} \left(\frac{1}{M_{H_1}^2}-\frac{1}{M_{H_2}^2}\right)^2,
\nonumber \\
&\simeq& 2.2 \times 10^{-45} {\rm cm}^2 \left( \frac{g_X s_\alpha c_\alpha}{10^{-2}} \right)^2 \left( \frac{75 \GeV}{M_{H_2}} \right)^4 \left( 1 - \frac{M_{H_2}^2}{M_{H_1}^2} \right)^2,
\end{eqnarray}
which can easily satisfy \texttt{LUX}~\cite{Akerib:2013tjd} limit for small mixing angle $\sin{\alpha}\lesssim 0.05$. This also means that direct detection can place a much stronger limit on $\sin\alpha$ for light dark higgs. Since the production cross section of this second higgs goes like $\sin^2{\alpha}$, it would be very challenge for probe it at the LHC. 

\section{Summary}\label{sec:summary}

In the letter, we have explored a possibility that the GeV scale $\gamma$-ray excess from the galactic center
is due to DM pair annihilation into a pair of dark Higgs, followed by the dark Higgs decay into the SM particles
through its small mixing with the SM Higgs boson.  Including the correlations among different parameters and 
varying $M_X$ and $M_{H_2}$ independently,   we find that the best fit is obtained if 
\[
M_X\simeq 95.0\GeV, \ \ \ M_{H_2}\simeq 86.7\GeV, \ \ \ 
\langle \sigma v\rangle\simeq 4.0 \times  10^{-26}\textrm{cm}^3\textrm{/s}
\]
If we impose $M_X \simeq M_{H_2}$, we get a similar result (see Table I). 
This information could be important inputs in dark matter models with dark Higgs boson.

At this stage we cannot make any strong statement about the existence of dark Higgs with mass close
to the DM mass $\sim 95$GeV.  However dark Higgs is very generic in DM models where 
DM is stabilized by some spontaneously broken local (or even global) dark gauge symmetries 
\cite{Ko:2014gha,Boehm:2014bia,Ko:2014loa,Baek:2011aa,Baek:2012uj,Baek:2012se,
Baek:2013qwa,Baek:2013dwa,Ko:2014nha,Ko:2014bka,Baek:2014jga,Khoze:2014woa,
Chen:2015nea}.  
Since the dark Higgs boson is  a SM singlet scalar, it is very difficult to find it at colliders 
although there are some interesting constraints from the LHC data on the SM Higgs signal strengths~\cite{Cheung:2015dta}.  
It is simply more difficult to produce them at colliders  when the mixing angle is small.  
It is very amusing to notice that indirect DM detection experiments can be  more sensitive 
to such a dark Higgs than the collider experiments. Compared with more popular dark 
photon scenario, it is more natural to have flavor dependent couplings of dark Higgs boson to the SM fermions,
since its couplings are basically the same as those of the SM Higgs boson modulo the mixing angle effect. 
This fact makes much easier DM model building \cite{Ko:2014gha,Boehm:2014bia}. 
It remains to be seen whether this fit  survives in the future data sets, and if there would be any indication 
of such a dark Higgs from the future collider  experiments. 

\begin{acknowledgments}
We are grateful to C.~Weniger for sharing the covariant matrix and an example script. We thank the KIAS 
Center for Advanced Computation for providing computing resources. YT thanks Xiaodong Li for helpful discussions. This work is supported in part by 
National Research Foundation of Korea (NRF) Research Grant 2012R1A2A1A01006053, 
and by the NRF grant funded by the Korea  government (MSIP) (No. 2009-0083526) through  
Korea Neutrino Research Center  at Seoul National University (PK). 
\end{acknowledgments}

\end{document}